\documentclass[12pt]{iopart}
\usepackage[numbers,square]{natbib}
\usepackage{booktabs}
\usepackage[colorlinks = true,
            linkcolor = cyan,
            urlcolor  = blue,
            citecolor = blue,
            anchorcolor = black]{hyperref}	
\usepackage{soul} 
\usepackage{setspace}
\usepackage{graphicx}
\usepackage{caption}

\usepackage[dvipsnames]{xcolor}

\usepackage{soul}

\makeatletter
\newcommand{\smallsym}[2]{#1{\mathpalette\make@small@sym{#2}}}
\newcommand{\make@small@sym}[2]{%
  \vcenter{\hbox{$\m@th\downgrade@style#1#2$}}%
}
\newcommand{\downgrade@style}[1]{%
  \ifx#1\displaystyle\scriptstyle\else
    \ifx#1\textstyle\scriptstyle\else
      \scriptscriptstyle
  \fi\fi
}
\makeatother

\begin{document}

\title{MHD Simulations on Magnetic Compression of Field Reversed Configurations}

\author{Yiming Ma$^1$, Ping Zhu$^{1,2*}$, Bo Rao$^{1*}$ and Haolong Li$^3$}

\address{
$1$ International Joint Research Laboratory of Magnetic Confinement Fusion and Plasma Physics, State Key Laboratory of Advanced Electromagnetic Engineering and Technology, School of Electrical and Electronic Engineering, Huazhong University of Science and Technology, Wuhan, Hubei 430074, China. \\
$2$ Department of Engineering Physics, University of Wisconsin-Madison, Madison, Wisconsin 53706, USA \\
$3$ College of Physics and Optoelectronic Engineering, Shenzhen University, Shenzhen 518060, China

}
\ead{zhup@hust.edu.cn, borao@hust.edu.cn}

\vspace{10pt}


\begin{abstract}

The magnetic compression has long been proposed a promising method for the plasma heating in a field reversed configuration (FRC), however, it remains a challenge to fully understand the physical mechanisms underlying the compression process, due to its highly dynamic nature beyond the one-dimensional (1D) adiabatic theory model [R. L. Spencer \textit{et al.}, Phys. Fluids \textbf{26}, 1564 (1983)]. In this work, magnetohydrodynamics (MHD) simulations on the magnetic compression of FRCs using the NIMROD code [C. R. Sovinec \textit{et al.}, J. Comput. Phys. \textbf{195}, 355 (2004)] and their comparisons with the 1D theory have been performed. The effects of the assumptions of the theory on the compression process have been explored, and the detailed profiles of the FRC during compression have been investigated. The pressure evolution agrees with the theoretical prediction under various initial conditions. The axial contraction of the FRC can be affected by the initial density profile and the ramping rate of the compression magnetic field, but the theoretical predictions on the FRC's length in general and the relation $r_s=\sqrt{2}r_o$ in particular hold approximately well during the whole compression process, where $r_s$ is the major radius of FRC separatrix and $r_o$ is that of the magnetic axis. The evolutions of the density and temperature can be affected significantly by the initial equilibrium profile and the ramping rate of the compression magnetic field. During the compression, the major radius of the FRC is another parameter that is susceptible to the ramping rate of the compression field. Basically, for the same magnetic compression ratio, the peak density is higher and the FRC's radius $r_s$ is smaller than the theoretical predictions.


\end{abstract}

\vspace{0.5pc}

\section{Introduction}

The field reversed configuration (FRC) is a compact toroid that is dominated by the poloidal magnetic field~\cite{tuszewski1988field}. The advantages of FRC as a magnetic confinement of high-temperature plasma include the high averaged beta~\cite{lin2017field}, the linear device geometry, the stability exceeding expectations from the magnetohydrodynamics (MHD) theory~\cite{steinhauer2011review, schwarzmeier1983magnetohydrodynamic}, and the natural divertor structure~\cite{steinhauer1996frc}. Besides these, an FRC plasma can maintain its magnetic structure during the supersonic translation~\cite{sekiguchi2018super, kobayashi2021experimental} and survive the violent collision and merging process~\cite{asai2021observation, asai2019collisional}. FRC is considered a promising candidate for compact fusion reactor due to its simplicity, high power density, and robustness~\cite{dettrick2021simulation}. To achieve fusion through the path of FRC, one approach is to maintain a steady state FRC in the high performance regime~\cite{binderbauer2015high} using the advanced biasing control~\cite{schmitz2016control} and neutral-beam injections~\cite{tuszewski2012field, gota2019formation}, and another is to compress FRC in a short time~\cite{slough2016staged, bol1972adiabatic, intrator2004high,slough2011creation}. Recently, an FRC device based on the collision merging is under construction in Huazhong University of Science and Technology ("HFRC")~\cite{zhang2020design, peng2022simulation}, and the preliminary experiments on the magnetic compression of the merged FRC are planned.


The magnetic compression has long been explored as a potentially effective method for the heating of FRCs~\cite{steinhauer2011review}. The FRX-C/LSM experiments show that the plasma temperature increases about 4 times through magnetic compression~\cite{rej1992high}, and experiments on the IPA device show that the FRC can be compressed to $\mathrm{keV}$~\cite{slough2011creation}. However, the short time scale of the compression process (typically on the order of tens of microseconds) and the drastically dynamic behaviors of the FRC plasma make it difficult to develop effective diagnoses. Previously a one-dimensional (1D) adiabatic theory~\cite{spencer1983adiabatic} developed to predict the scalings of FRC compression has been found consistent with the radially averaged temperature and density in the initial stage of FRC compression in the FRX-C/LSM experiments~\cite{rej1992high}. Later a NIMROD simulation of the fast magnetic compression of FRC plasma shows that the rise of peak pressure follows the theory prediction within the initial $10\mathrm{\mu s}$ in general~\cite{woodruff2008adiabatic}.

Despite these earlier studies, the applicability of the theory remains to be more systematically examined in simulations. In particular, the 1D theory is based on several assumptions that may over-simplify the more realistic situations. For example, the theory assumes a quasi-static and highly elongated FRC during compression and temperature or density profile effects, as well as the curvature of field lines in a 2D magnetic configuration are neglected. The exact extent to which these assumptions are valid or not is yet to be further assessed.

In this work, we perform simulations on the magnetic compression of the FRC using the resistive MHD model implemented in the NIMROD code~\cite{sovinec2004nonlinear}. To make the numerical results comparable with the analytical model, the single-fluid model is adopted as in Spencer's theory~\cite{spencer1983adiabatic}, and the physical parameters are set as close to the analytical model as possible. Our simulation results are able to confirm some key scaling laws of the adiabatic compression from the 1D theory predictions, but also provide a more comprehensive view of the FRC magnetic compression process closer to realistic regimes. The plasma pressure and $r_s=\sqrt{2}r_o$ relation agree with the theoretical predictions well. Under various conditions, the axial contraction of the FRC may slightly differ from but is overall consistent with the theory. By contrast, the evolutions of the radius, temperature and density are condition-sensitive. In general the radius is smaller and the peak density is higher than the theoretical predictions.

The rest of paper is organized as follows. The analytical model and the numerical approach are introduced in Sec.~\ref{sec: Analytical model and numerical model}. The simulation results are presented in Sec.~\ref{sec: Results}. Starting with a representative simulation case (Sec.~\ref{sec: typical result}), the effects of the initial density profile (Sec.~\ref{sec: density profile}) and the effects of the ramping rate of the compression magnetic field (Sec.~\ref{sec: ramping rate}) are discussed. Finally, the discussion and conclusions are presented in Sec.~\ref{sec: Discussion and conclusions}.

\section{Analytical model and numerical model}
\label{sec: Analytical model and numerical model}
\subsection{Analytical model}
\label{sec: analytical model}

For the purpose of comparison, we briefly review the 1D adiabatic model of FRC compression developed by Spencer~\cite{spencer1983adiabatic}. Assuming the quasi-static condition, the FRC compression process is approximated as a sequence of equilibrium states. For a highly elongated FRC, its 2D equilibrium solution reduces to the 1D relation

\begin{equation}
p_m = p + \frac{B_z^2}{2\mu_0} = \frac{B_w^2}{2\mu_0}
\label{eqn:equilibrium}
\end{equation}
where $p_m$ is the maximum pressure and $B_w$ is the magnetic field outside the FRC plasma, and the cylindrical coordinate system $(r, \theta, z)$ is introduced. From the ansatz that the pressure $p$ is only a function of flux $\Psi=2\pi \int_0^r B_z r^{'}dr^{'}$ and $\Psi$ is an even function of $r^2-r_o^2$, the following relation between the major radius of separatrix $r_s$ and that of the magnetic axis $r_o$ can be derived from Eq.~(\ref{eqn:equilibrium})~\cite{armstrong1981field}

\begin{equation}
  \label{eqn: rsro}
    r_s = \sqrt{2} r_o
\end{equation}
Considering that in the axial direction there is no plasma at $z \gg 0$, and the conservation of magnetic flux at $z=0$ and $z \gg 0$, the volume averaged plasma beta - $\langle \beta \rangle$ can be derived from the axial and radial force balance~\cite{armstrong1981field}
\begin{equation}
    \langle \beta \rangle =\frac{1}{\pi r_s^2} \int_0^{r_s} \frac{p}{B_w^2/(2\mu_0)} 2\pi rdr =  1 - \frac{x_s^2}{2}
\end{equation}
where the $x_s=r_s/r_w$ is the ratio of the separatrix radius to the wall radius.
Under the quasi-static assumption, the plasma velocity $\vec{u}=0$, and the total energy within the separatrix is

\begin{equation}
    E = \int \left( \frac{p}{\gamma-1} + \frac{B^2}{2\mu_0} \right) dV
    \label{eqn:E}
\end{equation}
For the adiabatic compression, we have $dE=dW$, where $dW=F_r dr+F_z dz$ is the work directly acting on the separatrix surface. Expanding $dE=dW$ using Eqs. (\ref{eqn:equilibrium})-(\ref{eqn:E}), the adiabatic compression scaling laws can be derived~\cite{spencer1983adiabatic}, which apply to both magnetic and wall compressions~\cite{intrator2008adiabatic, intrator2008physics}. For the magnetic compression that is the concern in this article, the scaling laws are listed in Table.~\ref{tab: FRC scaling}, where the $n_m$ is the maximum number density, $T_m$ the maximum temperature, $l$ the length of the FRC, and $\epsilon \in (0,1)$ is a flux profile index where the $\epsilon=0.25$ is most frequently used. To be consistent with other usages of the adiabatic model~\cite{intrator2008physics, tuszewski1988semiempirical}, $\epsilon=0.25$ is adopted in this article.

\begin{table}
    \caption{Spencer's scaling laws~\cite{spencer1983adiabatic} for flux compression, $\epsilon > 0$}
    \label{tab: FRC scaling}
    \centering
     \begin{tabular}{p{4cm}p{4cm}}
       \toprule
       parameter  & scaling law \\
       \cmidrule(r){1-2}

       $B_w$ & $ x_s^{-(3+\epsilon)} $ \\

       $p_m$ & $ x_s^{-2(3+\epsilon)}$ \\

       $n_m$  & $ x_s^{\frac{ -6(3+\epsilon) }{ 5 }} \langle \beta \rangle^{\frac{ -2(1-\epsilon) }{ 5 }} $  \\

       $T_m$  & $ x_s^{\frac{-4(3+\epsilon)}{5}} \langle \beta \rangle^{ \frac{ 2(1-\epsilon) }{ 5 } } $ \\

       $l$  & $  x_s^{ \frac{ 2(4+3\epsilon) }{ 5 } }  \langle \beta \rangle^{ \frac{-(3+2\epsilon)}{5} } $ \\

       \bottomrule
     \end{tabular}
     \label{tab:table}
\end{table}

\subsection{Numerical model}
\label{sec: numerical model}

Simulations in this work have been performed using the NIMROD code~\cite{sovinec2004nonlinear}. The resistive single-fluid MHD equations solved in simulations are as follows

\begin{equation}
    \frac{\partial n}{\partial t} + \nabla \cdot (n\vec{u})= 0
  \end{equation}

  \begin{equation}
    \rho(\frac{\partial \vec{u}}{\partial t} + \vec{u}\cdot \nabla\vec{u} ) = \vec{J}\times \vec{B} - \nabla p - \nabla \cdot \stackrel{\leftrightarrow}{\Pi}
  \end{equation}

  \begin{equation}
  \frac{n}{\gamma-1} (\frac{\partial}{\partial t} + \vec{u}\cdot \nabla ) T = -p \nabla \cdot \vec{u} + Q
  \end{equation}

  \begin{equation}
    \frac{\partial \vec{B}}{\partial t} = - \nabla \times \vec{E}
  \end{equation}

  \begin{equation}
    \vec{E} = -\vec{u}\times \vec{B} + \eta \vec{J}
\end{equation}
where $n$ is the plasma number density, $\rho$ the mass density, $p$ the pressure, $T=T_i+T_e$ the total temperature, $\vec{u}$ the plasma velocity, $\vec{J}$ the current density, $\stackrel{\leftrightarrow}{\Pi}$ the viscosity tensor, $\vec{B}$ the magnetic field, $\vec{E}$ the electric field, $\eta$ the resistivity and $Q$ represents the resistive heating source. For comparison with Spencer's scaling laws, the ideal MHD model of FRC compression is approximated in the dissipationless limit of the above resistive MHD simulation model. In particular, the Spitzer's resistivity model is used
\begin{equation}
    \eta(t) = \eta_0 \left( \frac{T(t)}{T_0} \right)^{-\frac{3}{2}}
\end{equation}
where $T_0$ is a constant that is set to the maximum temperature at $t=0 \mathrm{\mu s}$, $\eta_0/\mu_0=1\mathrm{m^2/s}$, which is much smaller than the typical resistivity of FRCs~\cite{peng2022simulation, milroy2010extended, guo2008improved}. Although the ohmic heating effect is not considered in the analytical model, the ohmic heating energy induced in the simulations is two orders of magnitude smaller than the internal energy and thus negligible. A parallel viscosity of $10^4$ $\mathrm{m^2/s}$ is also used following previous FRC simulations~\cite{milroy2010extended, macnab2007hall}. These dissipation parameters are introduced to maintain the numerical stability of the simulations, where only the minimal range of values are kept to ensure the numerical convergence of the simulation results. The time step of simulations is set to $\Delta t=5\times10^{-9}\mathrm{s}$ to resolve the Alfv\'enic scale of the dynamic compression process.

To apply the self-consistent boundary conditions in the NIMROD simulations, the normal component of $\vec{B}$ and the tangential component of $\vec{E}$ are specified and other components of $\vec{B}$ and $\vec{E}$ are calculated self-consistently. For the magnetic compression simulations, both electric and magnetic boundary conditions are assumed axisymmetric, so that they are applied to the $N=0$ Fourier component only. Similar to the previous simulations~\cite{milroy2010extended}, we set the vector potential at boundary as
\begin{equation}
    \label{eqn: vecA}
    \vec{A} = A_{\theta} \hat{\theta} = r_w B_z f(z) g(t) \hat{\theta}
\end{equation}
\begin{equation}
    f(z)=\frac{1}{1+e^{(z_1-z)/\lambda}+e^{(z-z_2)/\lambda}}
\end{equation}
to capture the effects of compression field coils, where $r_w$ is the wall radius, $B_z$ is a multiplier to the magnitude of the compression magnetic field, $f(z)$ is a function that determines the axial range of external field, $\lambda$, $z_1$ and $z_2$ are shaping parameters for $f(z)$, and $g(t)$ is a ramping function. The induced electrical field $\vec{E}$ is
\begin{equation}
    \vec{E} = -\frac{\partial \vec{A}}{\partial t} = -r_w B_z f(z) \dot{g}(t) \hat{\theta}
\end{equation}
and the magnetic field can be calculated according to $\vec{B}=\nabla \times \vec{A}$. By specifying the $E_\theta$ and $B_r$ respectively, it is expected to have a compression magnetic field $B_z(z)$ ramping with time, which also satisfies the divergence-free condition at boundary.

\section{Simulation Results}
\label{sec: Results}

\subsection{The dynamic process of magnetic compression}
\label{sec: typical result}

\begin{figure*}[!htbp]
  \centering
  \includegraphics[width=0.8\textwidth]{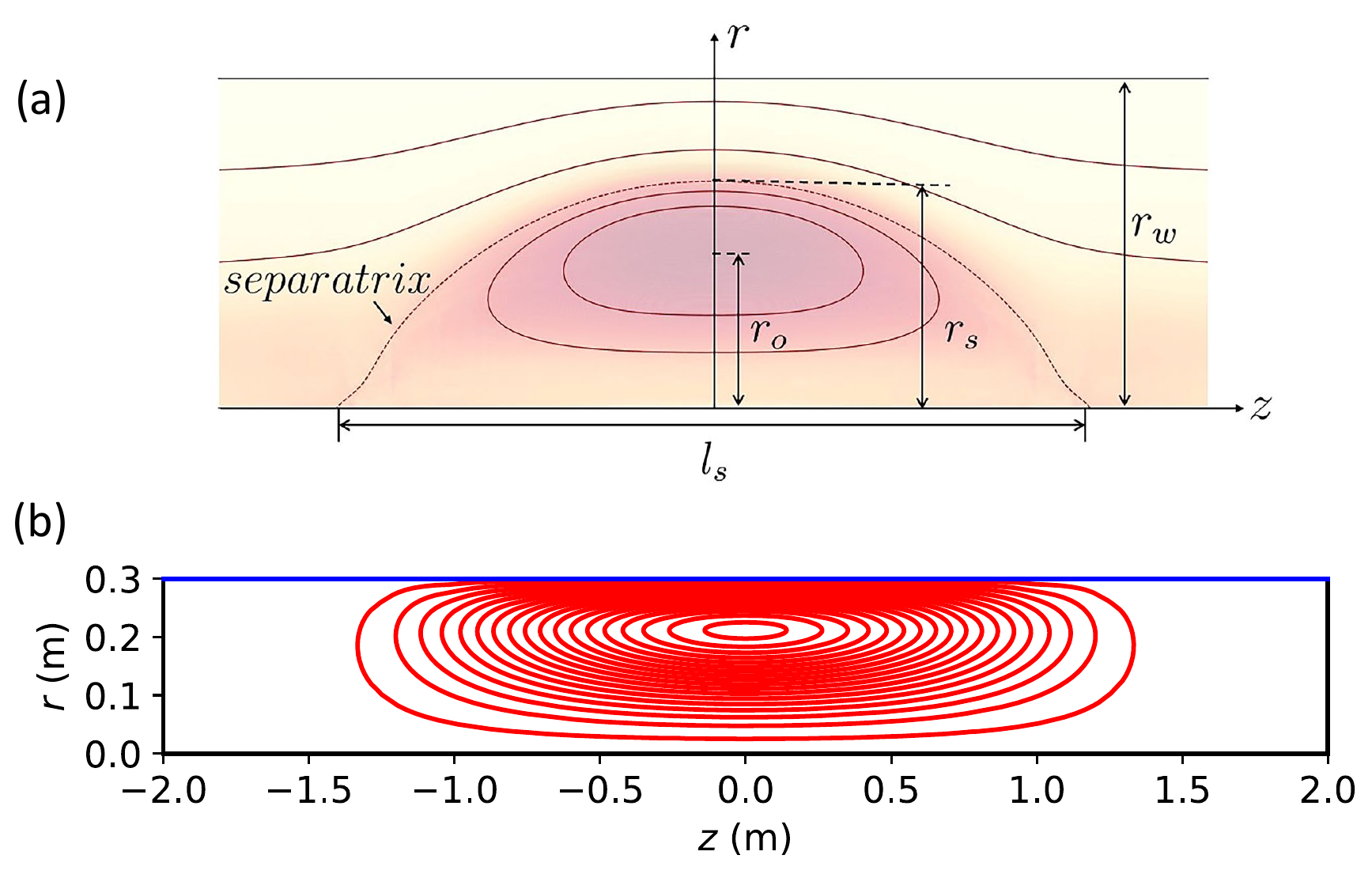}
  \caption{ (a) A schematic plot of the magnetic configuration (contour line) and plasma distribution (contour color) in the $r-z$ plane for an FRC. (b) Actual poloidal flux contour of the initial equilibrium (red) and the boundary at $r=r_w$ (blue)  for the externally applied compression magnetic field adopted in the simulations.}
  \label{fig: mesh_polfl}
\end{figure*}

The initial FRC equilibrium before the start of compression is prepared using the NIMEQ code~\cite{howell2014solving, li2021solving}. The axial length of the simulation domain is $4\mathrm{m}$ and the radius is $r_w=0.3\mathrm{m}$ as commonly designed for a modern FRC device. The grid has $16 \times 72$ 2D finite elements with the sixth-order polynomial basis functions along each direction in the poloidal plane. The plasma pressure outside the separatrix is assumed to be constant. The simulation starts from the simplest uniform number density profile, and the effects of the initial density profile will be discussed in Sec.~\ref{sec: density profile}. The initial magnetic field strength at wall $B_w$ is about $0.4\mathrm{T}$, density is $n_0=1\times10^{21}\mathrm{m^{-3}}$ and the maximum total temperature is $T_0=422\mathrm{eV}$.  The magnetic compression boundary condition discussed in Sec.~\ref{sec: numerical model} is applied from $z=-2\mathrm{m}$ to $z=2\mathrm{m}$ at $r=r_w$ as shown as Fig.~\ref{fig: mesh_polfl}, and all boundaries are assumed to be the no-flow solid walls.

\begin{figure*}[!htbp]
  \centering
  \includegraphics[width=1.0\textwidth]{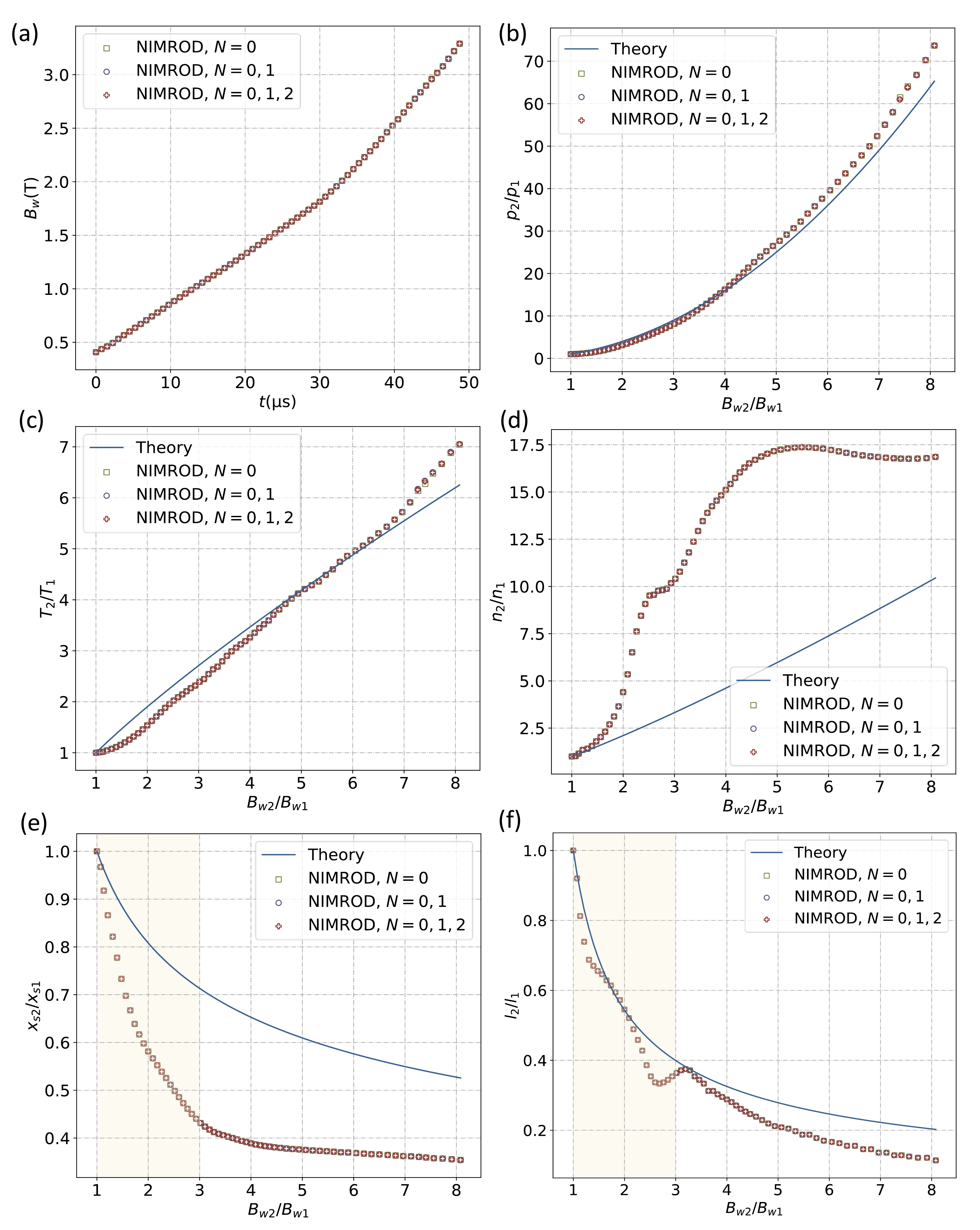}
  \caption{ Comparisons of the theory (blue lines) and representative simulation results with the toroidal mode number $N=0$ (green rectangles), $N=0,1$ (purple circles) and $N=0,1,2$ (red plus signs), and (a) the compression field $B_w$ variation with time, for the (b) pressure ratio $p_2/p_1$, (c) temperature ratio $T_2/T_1$, (d) density ratio $n_2/n_1$, (e) radius ratio $x_{s2}/x_{s1}$ and (f) length ratio $l_2/l_1$ as functions of the magnetic compression ratio $B_{w2}/B_{w1}$. The majority of the volume reduction occurs during the stage $B_{w2}/B_{w1} < 3$ (yellow shade). }
  \label{fig: typical results}
\end{figure*}

For a linear ramping of the applied magnetic field $B_w$ at wall, the consequent FRC evolution during the magnetic compression process is obtained from NIMROD simulations. The comparisons between the NIMROD results and the 1D theory are shown in Fig.~\ref{fig: typical results}, where the simulation cases are performed with different maximum toroidal mode number $N=0, 1$ and $2$ respectively without any initial perturbation internal to the simulation domain. As can be seen from Fig.~\ref{fig: typical results}, the simulation results with different maximum toroidal mode numbers are exactly the same, and no $N>0$ instability is observed.

For comparison with the 1D theory, the maximum pressure $p_m$, maximum total temperature $T_m$, maximum plasma number density $n_m$ and maximum $B_z$ at the $z=0$ plane are plotted in Fig.~\ref{fig: typical results}  respectively, where $x_s=r_s/r_w$ is the relative radius and $l$ is the length of the FRC. The subscript ``1" and subscript ``2" represent the initial equilibrium state at $t=0$ and the final compressed state at the corresponding time respectively. The curves of theoretical predictions are plotted using scaling laws shown in Table.~\ref{tab: FRC scaling}. The simulation results for all $p_m$, $T_m$ and $l$ are in good agreement with the theory during the compression. The FRC shrinks at a faster rate in the radial direction and the maximum number density at the middle plane rises more rapidly compared to the theoretical predictions, which are consistent with the experimental results~\cite{rej1992high}. Close inspection of the Fig.~\ref{fig: typical results} shows that the majority of the volume reduction occurs during $B_{w2}/B_{w1}<3$, after which the rates of shrinkage in both the axial and radial directions decrease. The simulation runs until the magnetic compression ratio $B_{w2}/B_{w1}\sim 8$. This is because, on the one hand, at the late stage of the compression process the length of the FRC shrinks to less than $0.3\mathrm{m}$ when the single-fluid MHD model becomes less applicable since the finite Larmor radius and kinetic effects may no longer be negligible; on the other hand, the mesh does not get finer accordingly as the FRC shrinks, so that the numerical error becomes more significant. The agreement between the simulation results and the theory for the curves of $p_m$, $T_m$ and $l$ exceeds the expectation, because the FRC is quickly compressed from elongated to oblate through a quite dynamic process in the simulation far different from the quasi-static condition assumed in the theory. The comparison suggests that the theory appears applicable beyond the limitation from its own assumptions. The applicability of the theory will be further checked in Sec.~\ref{sec: density profile} and Sec.~\ref{sec: ramping rate}.

\begin{figure*}[!htbp]
  \centering
  \includegraphics[width=0.9\textwidth]{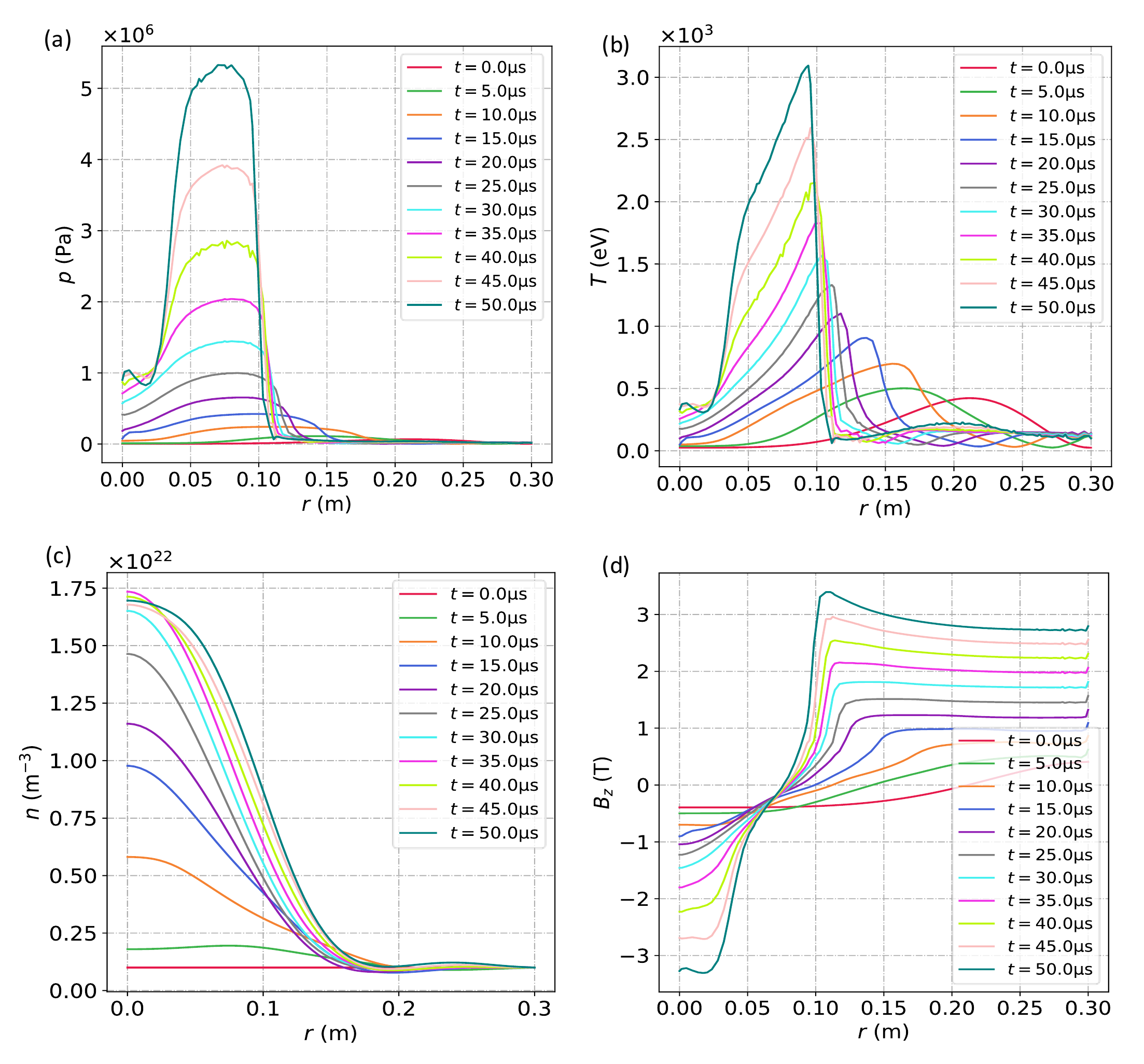}
  \caption{Radial profiles along $z=0$ of the (a) pressure, (b) temperature, (c) density and (d) $B_z$ at different times during the compression process respectively. The simulation is same as the $N=0$ case shown in Fig.~\ref{fig: typical results}.}
  \label{fig: typical 1d}
\end{figure*}
\begin{figure*}[!htbp]
  \centering
  \includegraphics[width=0.6\textwidth]{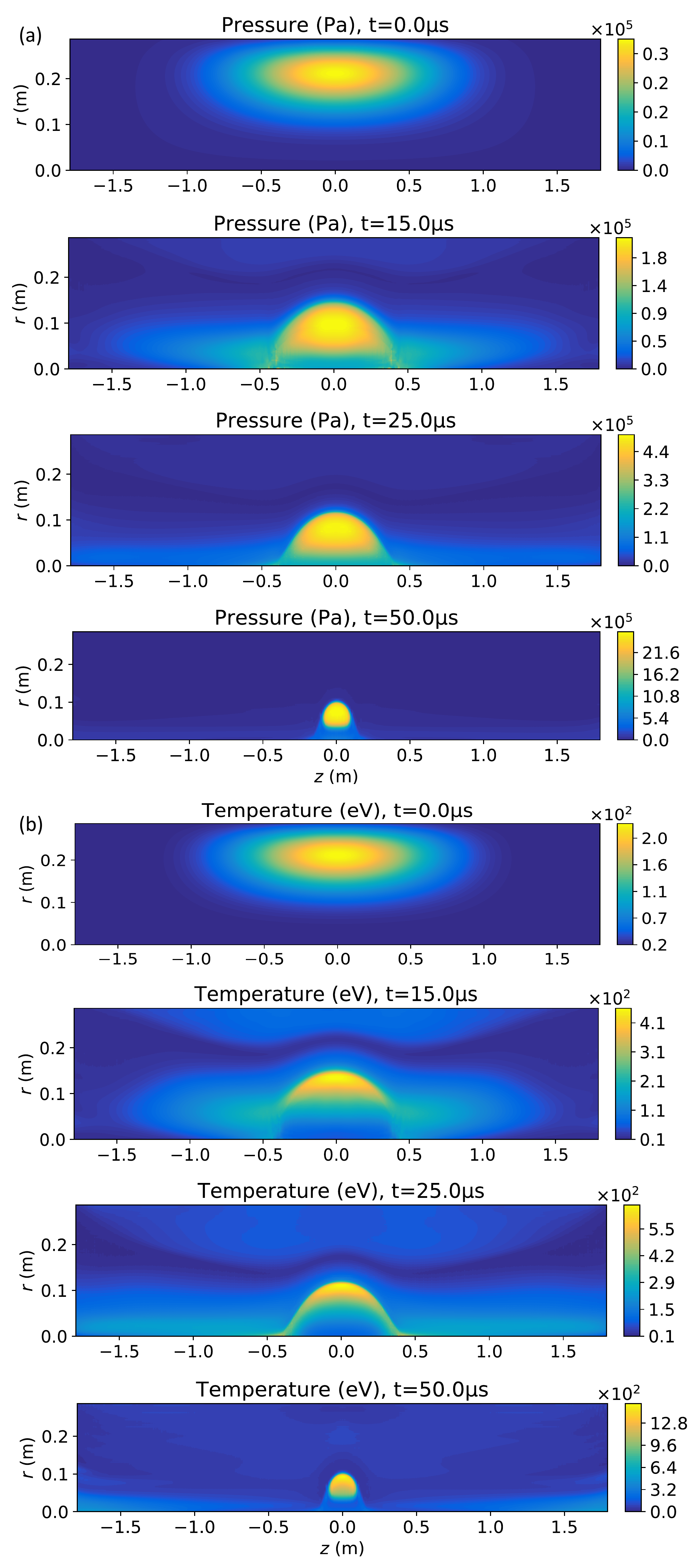}
  \caption{(a) Pressure and (b) temperature contours in the $r-z$ plane of the same simulations as shown in Fig.~\ref{fig: typical results} at $t=0\mathrm{\mu s}$, $t=15\mathrm{\mu s}$, $t=25\mathrm{\mu s}$, and $t=50\mathrm{\mu s}$ respectively. }
  \label{fig: typical contour}
\end{figure*}

The radial profiles of the pressure, temperature, density and the axial magnetic field along the middle plane $z=0$ at different times can be used to illustrate the compression process (Fig.~\ref{fig: typical 1d}). The radii of zero crossings of the $B_z$ shown in Fig.~\ref{fig: typical 1d} (d) are those of the magnetic axis or O-point of FRC, i.e. $r_o$ at corresponding times. It is noted that the pressure and the temperature profiles continue to steepen on both sides of $r_o$, and form a narrow region of sharp gradients by the last stage of the compression process. Furthermore, the temperature outside $r_o$ is higher than the inside instead of peaking around $r_o$ as in the case of pressure, which can also be seen from the contours of the pressure and temperature in Fig.~\ref{fig: typical contour}. The possible reason for forming this kind of pressure and temperature profiles is that the FRC compression rate is slower than the maximum acoustic speed, while the thermal conduction terms especially the parallel thermal conduction are not taken into account in the adiabatic model for pressure in simulations. The pressure and temperature contours in the poloidal plane at a sequence of time show the expected FRC shrinking toward the center and the transition from elongated to oblate shape during the compression process (Fig.~\ref{fig: typical contour}). Moreover, the FRC maintains a good symmetry about the middle plane at $z = 0$ in the poloidal plane throughout the compression process, and no $N=0$ instability such as the Roman candle instability~\cite{tuszewski1988field} occurs. In contrast the radial density profile loses its initial hollow structure after $t=10\mathrm{\mu s}$, and the maximum density accumulates at the cylinder axis at $r=0$ (Fig.~\ref{fig: typical 1d}), which however does not increase monotonically with time but decreases slightly after $t=35\mathrm{\mu s}$. The maximum values of pressure, temperature and density are not at the same location as shown in Fig.~\ref{fig: typical 1d}, which may be why they could be higher than the theoretical predictions at certain given time (Fig.~\ref{fig: typical results}).

\begin{figure*}[!htbp]
  \centering
  \includegraphics[width=0.7\textwidth]{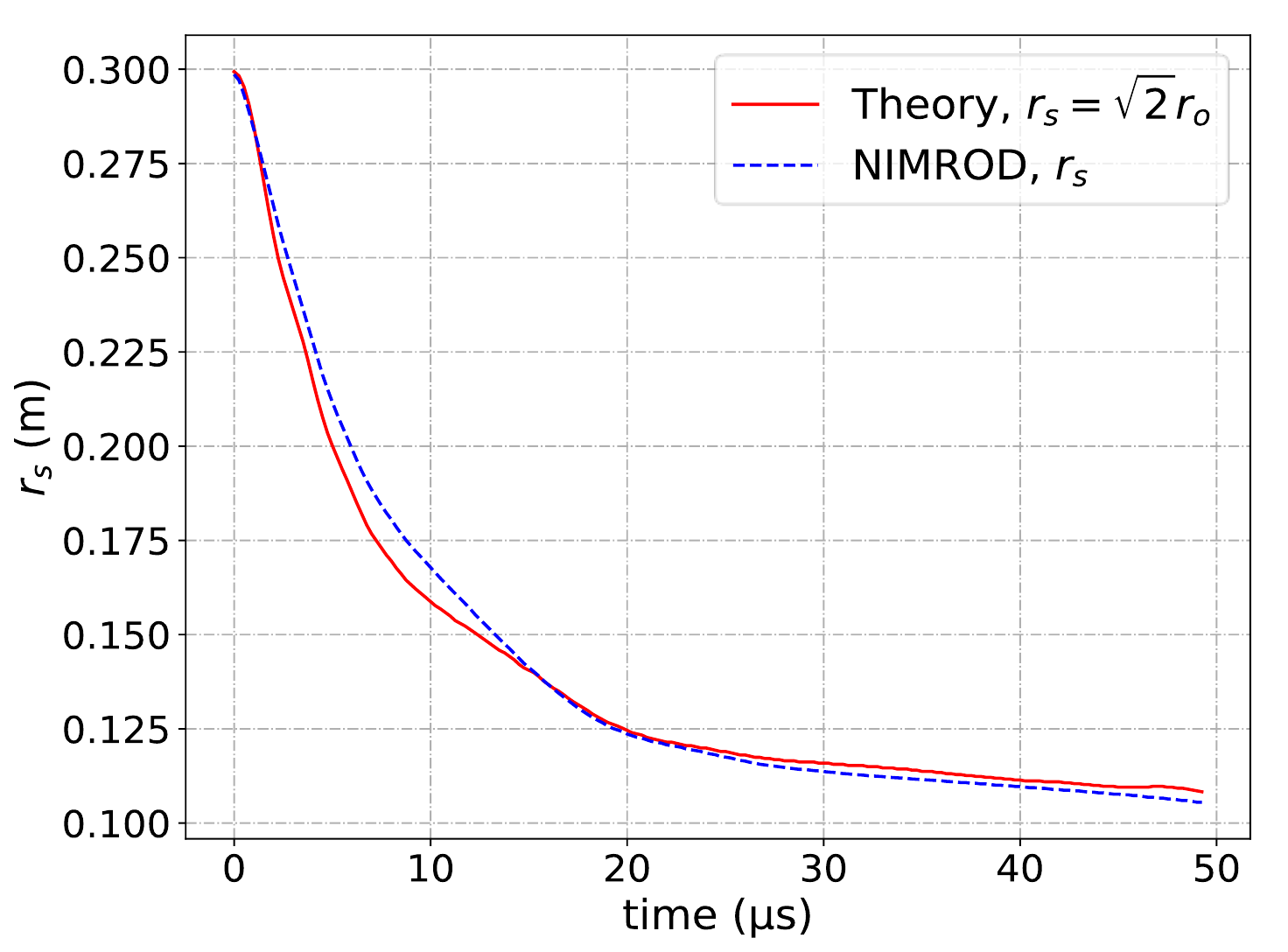}
  \caption{Time history of $r_s$ from the theoretical prediction (red solid line) and the same simulations as shown in Fig.~\ref{fig: typical results}  (blue dashed line).}
  \label{fig: rors}
\end{figure*}

For all the simulations in absence of initial perturbation, not only that no MHD instability is observed, but also that the magnetic field structure in the radial direction remains essentially same when higher toroidal components (N=1, 2) are included. The $r_s$ and the $\sqrt{2}r_o$ are approximately equal during the compression process, where the major radius of the magnetic axis $r_o$ is the radius of the zero crossing of the $B_z$ at $z=0$, and the radius of the separatrix $r_s$ is the position where the flux $\Psi=0$ outside the $r_o$ at $z=0$ (Fig.~\ref{fig: rors}), which confirms the theory prediction in Eq.~(\ref{eqn: rsro}) that $r_s=\sqrt{2}r_o$.

\subsection{Effects of the initial number density profile}
\label{sec: density profile}

\begin{figure*}[!htbp]
  \centering
  \includegraphics[width=0.7\textwidth]{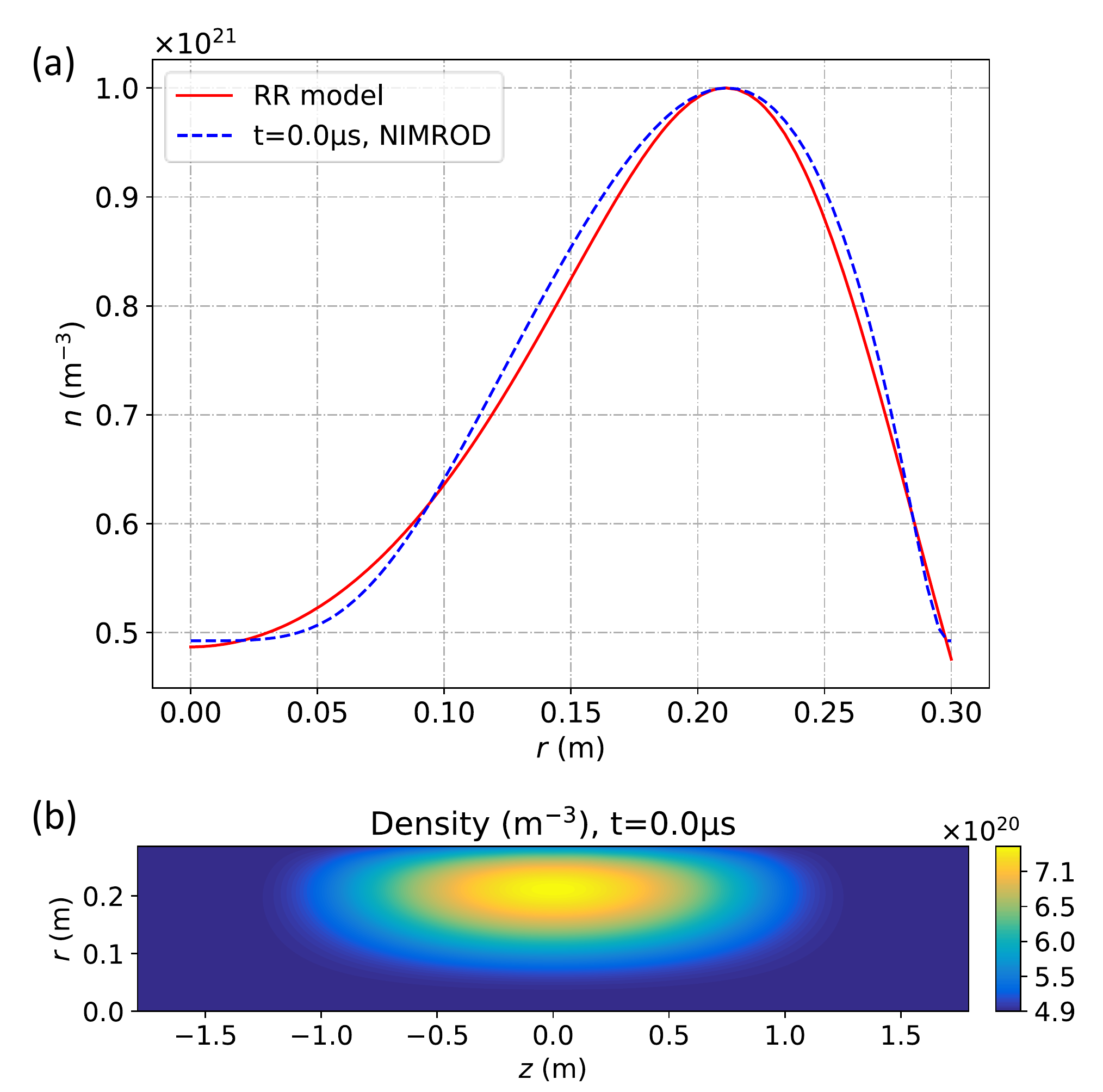}
  \caption{(a) The 1D density profile along $z=0$ of the initial equilibrium (blue dashed line) and the RR model where the shape parameter $k=0.9$ (red solid line). (b) The 2D density contour of the initial equilibrium.}
  \label{fig: neq1d2d}
\end{figure*}

The representative simulation in Sec.~\ref{sec: typical result} starts with the simplest uniform number density profile. In reality the measured number density profile of FRC equilibrium is usually nonuniform such as that in the rigid rotor (RR) model~\cite{renneke2007global, okada1989reduction}. In order to study the effects of number density profile on the compression process, a nonuniform initial number density distribution which is consistent with the RR profile at the $z=0$ plane is adopted as follows
\begin{equation}
  n(\Psi) = n_0 \left( \frac{p(\Psi)}{p_0} \right)^{\Gamma}
\end{equation}
where $\Psi$ is the poloidal flux, $n_0=1\times10^{21}\mathrm{m^{-3}}$ is the initial density used in Sec.~\ref{sec: typical result}, $p_0$ is the pressure at the magnetic axis and $\Gamma$ is the profile index where $\Gamma=0$ implies a uniform initial density distribution as the case in Sec.~\ref{sec: typical result}. For $\Gamma=\frac{1}{4}$ used in this section, the following radial density profile from the RR model with $k=0.9$ at the $z=0$ plane~\cite{steinhauer2011review, armstrong1981field}
\begin{equation}
  \label{eqn: RR profile}
  n(r) = n_m \mathrm{sech}^2 \Bigg[ k \bigg(\frac{2r^2}{r_s^2}-1 \bigg) \Bigg]
\end{equation}
can be essentially reproduced, where $k$ is the shape parameter and $n_m$ is the maximum density. Similar number density profile is also measured in a typical shot on FRX-L~\cite{renneke2007global}. The specific initial 1D profile and 2D distribution of number density are shown in the Fig.~\ref{fig: neq1d2d}. Compared with the simulation in Sec.~\ref{sec: typical result}, the equilibrium pressure profile and boundary conditions remain the same, and only the initial density and temperature distributions are changed.

\begin{figure*}[!htbp]
  \centering
  \includegraphics[width=1.0\textwidth]{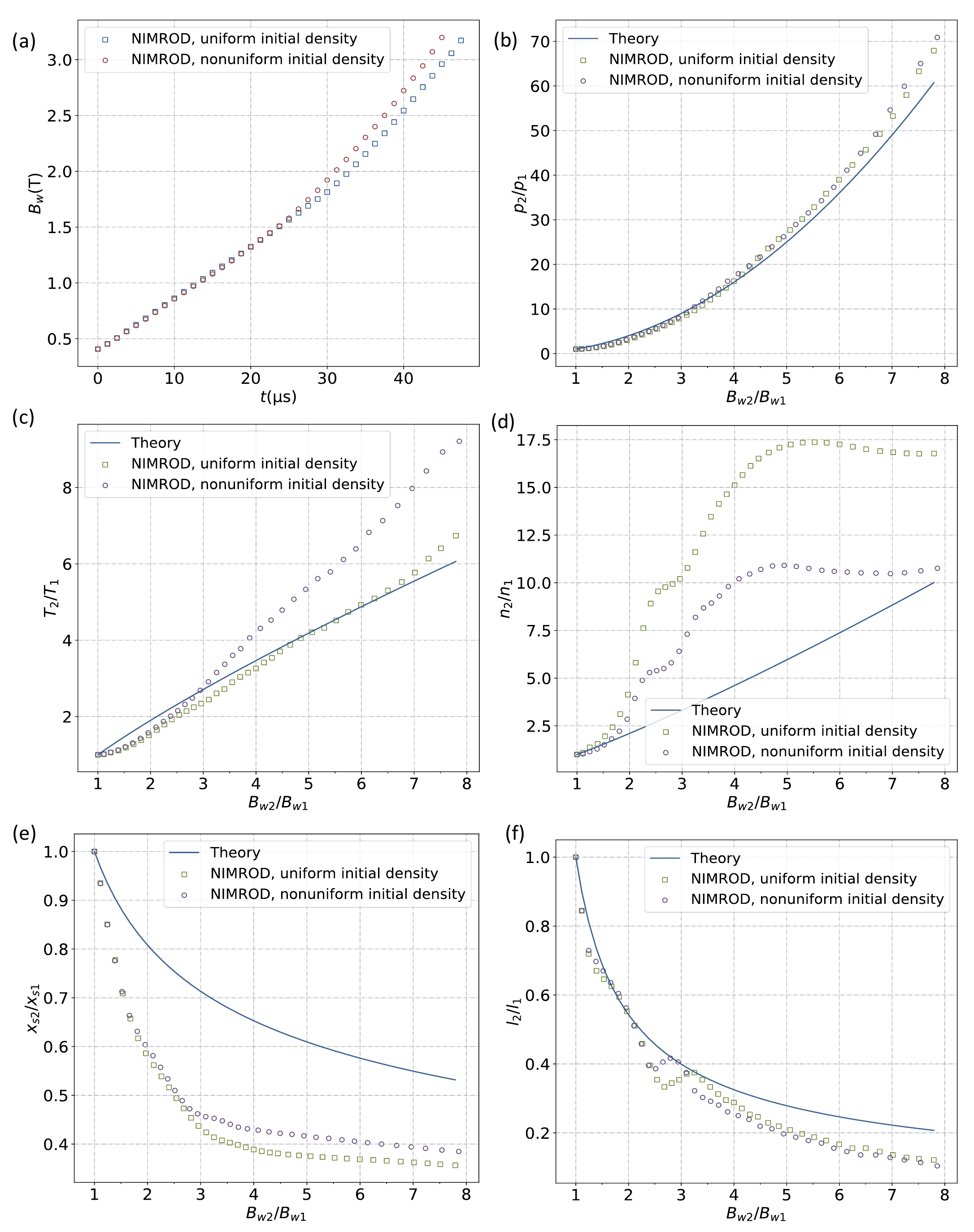}
  \caption{Simulation results with nonuniform initial density (purple circles) in comparisons with the 1D theory (blue lines) and the simulation results with the uniform initial density that is same as the $N=0$ case shown in Fig.~\ref{fig: typical results} (green rectangles). (a) The time dependence of the compression field $B_w$. (b) The pressure ratio $p_2/p_1$, (c) temperature ratio $T_2/T_1$, (d) density ratio $n_2/n_1$, (e) radius ratio $x_{s2}/x_{s1}$, and (f) length ratio $l_2/l_1$ as functions of the magnetic compression ratio $B_{w2}/B_{w1}$ for different initial density profiles. }
  \label{fig: eff den}
\end{figure*}

\begin{figure*}[!htbp]
  \centering
  \includegraphics[width=0.7\textwidth]{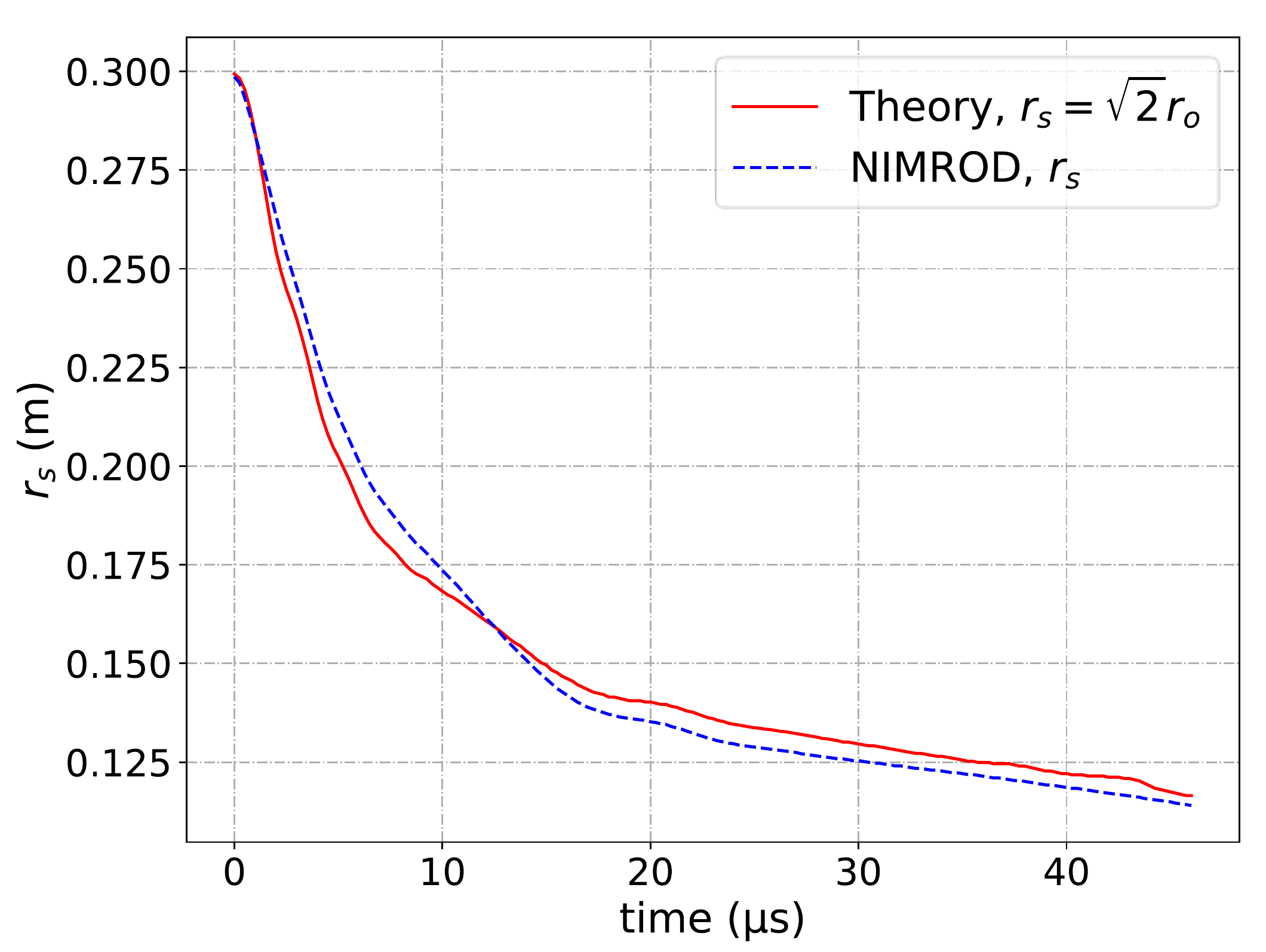}
  \caption{The time history of the $r_s$ of the theoretical prediction (red solid line) and the same simulation as shown in Fig.~\ref{fig: eff den}  (blue dashed line).}
  \label{fig: rors nonden}
\end{figure*}

\begin{figure*}[!htbp]
  \centering
  \includegraphics[width=0.7\textwidth]{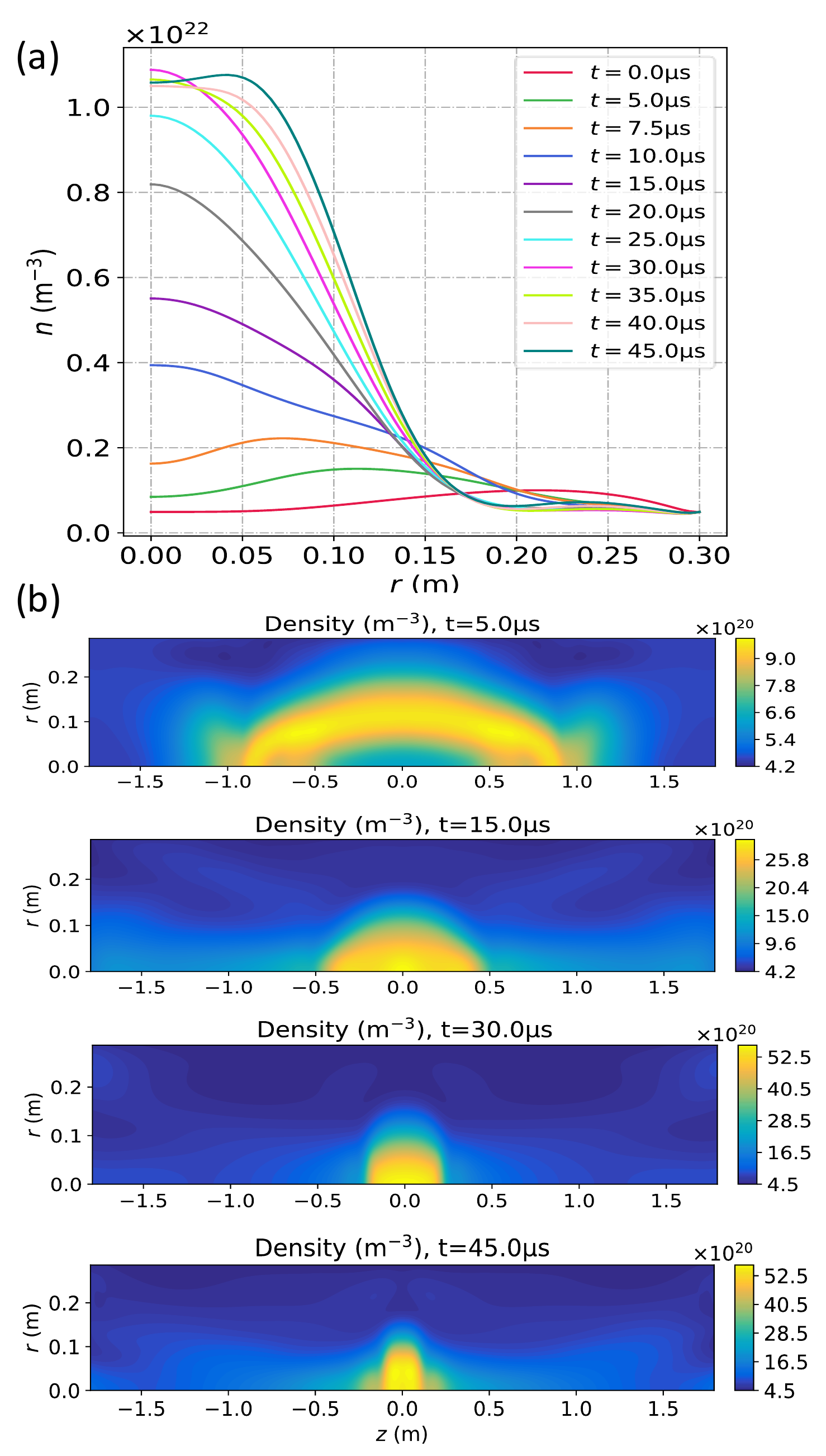}
  \caption{(a) The 1D radial density profile along $z=0$ at different times during the compression. (b) The density contours at different times during the compression. The simulation is same as the case shown in Fig.~\ref{fig: eff den}.}
  \label{fig: dens evo nonu}
\end{figure*}

Fig.~\ref{fig: eff den} shows the simulation results with nonuniform initial density compared with the 1D theory and the simulation results with the uniform initial density. The simulation results with the uniform initial density are same as the case shown in Fig.~\ref{fig: typical results}. Because the $B_w$ is the sum of the magnetic field externally applied at the boundary condition and induced by the plasma, the $B_w$ of the two cases shown in Fig.~\ref{fig: eff den}(a) are slightly different after $t>25\mathrm{\mu s}$, even though the externally applied components are the same. As shown in Fig.~\ref{fig: eff den}, the peak pressure ratio $p_2/p_1$ curves of the two  simulation cases are basically identical. The $l_2/l_1$ curves of the two simulation cases differ slightly in the interval of $B_{w2}/B_{w1}>2.5$ and $B_{w2}/B_{w1}<4$, and the simulation case with nonuniform initial density shrinks slightly slower than the uniform initial density case in the radial direction after $B_{w2}/B_{w1}>3$. The evolutions of the temperature and density during the compression process are most affected by the initial density profile. Compared to the case with uniform initial density, the maximum temperature of the case with nonuniform initial density profile rises more rapidly after $B_{w2}/B_{w1} > 2.5$, and the maximum density rises more slowly after $B_{w2}/B_{w1} > 2$. Moreover, a close inspection of the density profile is shown in Fig.~\ref{fig: dens evo nonu}. Similar to the simulation results shown in Fig.~\ref{fig: typical 1d}, the radial density profile loses the hollow density structure before $t=10\mathrm{\mu s}$ during the compression process, and the density profile changes from elongated to oblate. Furthermore, as can be seen from the Fig.~\ref{fig: rors nonden}, the time history of $r_s$ from simulation confirms the theoretical prediction $r_s=\sqrt{2}r_o$, similar to the case with uniform initial density. Overall, these results suggest that the different initial density profiles mainly affect the evolutions of the temperature and density during the compression, and the radial density profile loses its hollow structure during the compression. A more peaked initial density profile basically leads to a higher maximum temperature and lower maximum density. Different initial density profiles slightly affect the FRC volume variation in time, but have nearly no effects on the pressure evolution or the relation between $r_s$ and $r_o$.

\subsection{ Ramping rate effects of compression magnetic field}
\label{sec: ramping rate}


The magnetic compression process of the FRC usually lasts only less than $70\mathrm{\mu s}$ in experiments and simulations~\cite{slough2011creation, rej1992high, woodruff2008adiabatic}, and the volume of the FRC shrinks drastically during the compression. One way to study the dynamic effects of the compression process, in contrast to the quasi-static condition assumed in the 1D theory, may be to vary the ramping rate of the externally applied compression magnetic field. In the approach of the NIMROD code, the different ramping rates are achieved by changing the multiplier $B_z$ in Eq.~(\ref{eqn: vecA}).

\begin{figure*}[!htbp]
  \centering
  \includegraphics[width=1.0\textwidth]{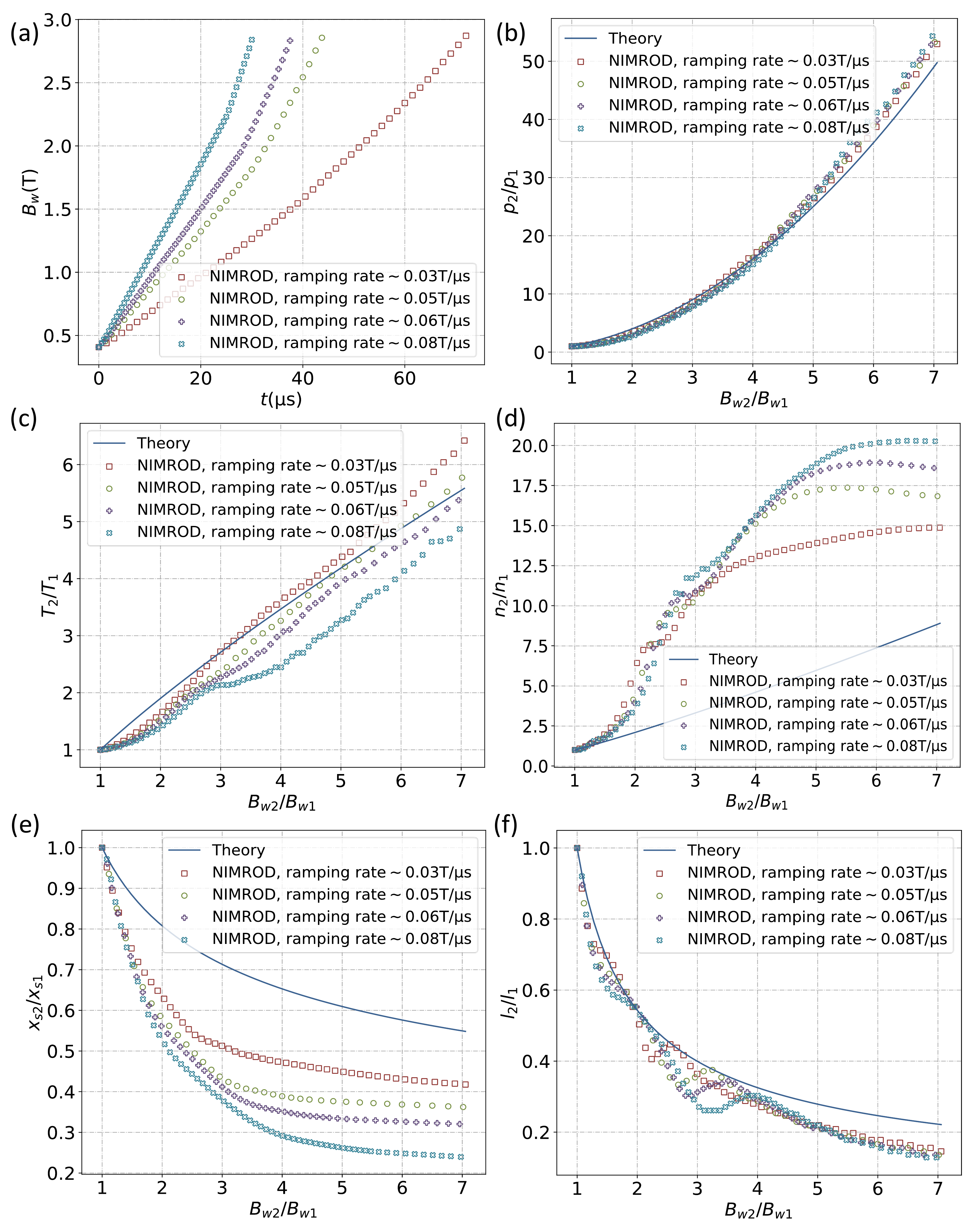}
  \caption{The comparison of the 1D theory (blue lines) and four simulation cases with different ramping rates of the compression magnetic field. The simulation with the ramping rate $\sim 0.05\mathrm{T/\mu s}$ is same as the $N=0$ case shown in Fig.~\ref{fig: typical results}. (a) The compression field $B_w$ as a function of time. (b) The pressure ratio $p_2/p_1$, (c) temperature ratio $T_2/T_1$, (d) density ratio $n_2/n_1$, (e) radius ratio $x_{s2}/x_{s1}$, and (f) length ratio $l_2/l_1$ as functions of the magnetic compression ratio $B_{w2}/B_{w1}$ for different ramping rates of the $B_w$.}
  \label{fig: ramping rate}
\end{figure*}

Simulation cases with four different ramping rates of the compression magnetic field are compared (Fig.~\ref{fig: ramping rate}). The simulation case with an approximate ramping rate of $0.05\mathrm{T/\mu s}$ (green circles in Fig.~\ref{fig: ramping rate}) is same as the one presented in Sec.~\ref{sec: typical result}. The ramping rates of other three cases are approximated as $0.03\mathrm{T/\mu s}$, $0.06\mathrm{T/\mu s}$ and $0.08\mathrm{T/\mu s}$ respectively. The increment of $B_w$ is approximately but not exactly linear with time, for the reason explained earlier that it includes both the externally applied boundary and the plasma induced contributions. A faster ramping rate of the $B_w$ basically implies a more intense dynamic process. As the ramping rate of the compression field increases, the peak temperature rises more slowly, the peak density rises more rapidly after $B_{w2}/B_{w1}> 3$, and the radius of the FRC decreases faster with $B_{w2}/B_{w1}$. The difference in ramping rate has a slight effect on the axial length variation in the range $2<B_{w2}/B_{w1}<4$, but has no effects on the pressure evolution. The 1D theory prediction for the pressure evolution remains in good agreement with the simulation results despite the dynamic effects considered in the simulation.

\section{Discussion and conclusions}
\label{sec: Discussion and conclusions}

In summary, the simulations on the magnetic compression of FRCs using the NIMROD code, and their comparisons with the Spencer's adiabatic theory~\cite{spencer1983adiabatic} have been presented in this article. The single-fluid MHD model is adopted, and the physical parameters are set as close proxies to the analytical model for the purpose of comparison. A set of self-consistent boundary conditions have been implemented to model the effects of the externally applied magnetic field for compression. The simulation results for the pressure evolution agree well with the theory prediction for various initial conditions. The axial contraction of the FRC is slightly faster than the theoretical prediction after $B_{w2}/B_{w1}>3$. In the range $B_{w2}/B_{w1}>2$ to $B_{w2}/B_{w1}<4$, the evolution of the axial length is somewhat influenced by the initial density profile and the ramping rate of the compression field. Nevertheless, the theoretical prediction on the FRC's length and the relation $r_s=\sqrt{2}r_o$ hold approximately well during the entire compression process.

The initial density profile of the FRC and the ramping rate of the compression magnetic field both have significant effects on the evolutions of the temperature and density. Compared to the case with uniform initial density, for the case with nonuniform initial density profile, the peak density rises more slowly after $B_{w2}/B_{w1}>2$, and the peak temperature rises more rapidly after $B_{w2}/B_{w1}>2.5$. The major radius of the FRC is another parameter that is susceptible to the ramping rate of the compression field. A faster ramping rate of the compression field leads to a faster decrease in FRC's radius. In general, for the same magnetic compression ratio, the peak density is higher and the radius is smaller than the theoretical predictions, which are consistent with the experimental results~\cite{rej1992high}.

During the compression process, the pressure and temperature maintain a hollow structure, while their gradients increase on both radial sides of the magnetic axis at $r_o$. Moreover, the temperature does not have a peaked profile near the $r_o$ as the pressure, and the temperature outside $r_o$ is higher than the inside. Unlike the pressure and temperature profiles, the radial density profile loses its hollow structure during the compression.

Given the confidence in the simulation results in the ideal-like MHD regime, the simulations can be extended to the more experimentally relevant regimes and scenarios. For examples, the finite-Larmor radius and the two-fluid effects can be included, along with more realistic values of resistivity and anisotropic thermal conductivities. When initial and dynamic perturbations are considered, MHD instabilities may play an important role. We plan on addressing these issues in future work.

\section{Acknowledgement}

The authors thank Profs. C. R. Sovinec, B. H. Deng and Z. J. Wang for helpful discussions. This work was supported by the National Key Research and Development Program of China (Grant No. 2017YFE0301805 and Grant No. 2019YFE03050004), the Fundamental Research Funds for the Central Universities at Huazhong University of Science and Technology (Grant No. 2019kfyXJJS193), the National Natural Science Foundation of China (Grant No. 51821005), and the U.S. Department of Energy (Grant Nos. DE-FG02-86ER53218 and DE-SC0018001). The computing work in this paper was supported by the Public Service Platform of High Performance Computing by Network and Computing Center of HUST. The authors  are very grateful for the supports from the NIMROD team and the J-TEXT team.

\normalsize
\bibliography{references}
\clearpage


\clearpage

\clearpage

\clearpage

\clearpage

\clearpage

\clearpage

\clearpage

\clearpage

\clearpage

\clearpage

\end{document}